\documentclass[%
 amsmath,amssymb,
 aps, prx, twocolumn, superscriptaddress,
 10pt
]{revtex4-2}
\usepackage{comment}
\usepackage{graphicx}% Include figure files
\usepackage{dcolumn}% Align table columns on decimal point
\usepackage[dvipsnames]{xcolor} % Make sure this line is present
\usepackage{bm}% bold math
\usepackage{braket}
\usepackage{soul}

\graphicspath{ {./figs/} }

\newcommand{\EqRef}[1]{Eq.~\eqref{#1}}
\newcommand{\FigRef}[1]{Fig.~\ref{#1}}

\begin{document}

% \preprint{APS/123-QED}

\title{Moment closure through spectral expansion in open stochastic systems}

\author{Gianni V. Vinci}
\email{gianni.vinci@iss.it}
\affiliation{Natl. Center for Radiation Protection and Computational Physics, Istituto Superiore di Sanità, 00161 Roma, Italy}
\affiliation{PhD Program in Physics, “Tor Vergata” University of Rome, 00133 Roma, Italy}

\author{Roberto Benzi}
\affiliation{Sino-Europe Complex Science Center, School of Mathematics
North University of China, Shanxi, 030051 Taiyuan, China}
\affiliation{Dept. of Physics and INFN, “Tor Vergata” University of Rome, 00133 Roma, Italy}

\author{Maurizio Mattia}
\affiliation{Natl. Center for Radiation Protection and Computational Physics, Istituto Superiore di Sanità, 00161 Roma, Italy}

\begin{abstract}
The derivation of dynamical laws for general observables (or moments) from the master equation for the probability distribution remains a challenging problem in statistical physics.
Here, we present an alternative formulation of the general spectral expansion, which clarifies the connection between the relaxation dynamics of arbitrary moments and the intrinsic time scales of the system.
Within this framework, we address the moment-closure problem in a way that streamline the conventional treatment of open systems.
The effectiveness of the theory is illustrated by deriving analytical expressions for two representative cases: spectral amplification in stochastic resonance and the moment dynamics of a non-Gaussian system, namely the Bessel process with constant drift.
We also identify a direct relationship between our theory and the Koopman operator approach. 
Finally, we apply our approach to the nonlinear and out-of-equilibrium mean-field description of interacting excitatory and inhibitory populations.
\end{abstract}

\maketitle


%\tableofcontents

Any dynamical system with or without noise can be studied by examining the continuity equation for the probability density $p(x,t)$ of finding the system in the state $x$ at time $t$.
This principle is at the core of statistical physics since the seminal work of Boltzmann on kinetic theory \cite{sethna2021statistical}.
Experimentally, the same systems are usually probed by measuring observables $m(t)$ that can be defined as time-dependent averages of functions of the state of the system over the time-dependent probability distribution.
One of the main goals has always been to reduce the dimensionality of the problem by finding self-consistent dynamical laws for a finite set of observables \cite{levermore1996}. 
We could argue that in model building, the most difficult challenge is to choose such an optimal set of variables, which always requires a deep understanding of the problem at hand \cite{onsager1953fluctuations,ma1985statistical}.
Unfortunately, in the general nonlinear case, the system cannot be closed directly in a straightforward way, and some assumptions of a different nature are required; this problem is typically referred to as moment closure \cite{kuehn2016}.

Recently, we introduced a possible solution to the moment-closure problem relying on the spectral decomposition of the continuity equation \cite{MattiaVinci2024}.
Here, we prove the general applicability of this spectral-based moment closure to a wide class of stochastic dynamical systems.
This hybrid approach, that we refer to as moment projection, has a number of advantages over standard spectral expansion, in particular when the system under examination is open, i.e., it has a time dependence through an external forcing.
This approach makes clear the relation between the time scales of the system and the relaxation dynamics of arbitrary observables.
Despite the fact that the approach here presented is formally equivalent to standard spectral expansion \cite{risken1996}, the latter is more involved in terms of analytical calculations.
In fact, most of the time, even when the eigenvalues and eigenfunctions are known, one must resort to numerical integration to find the coefficients of the expansion.
On the contrary, our formulation has a number of general constraints that allow greater analytical control.
We show the effectiveness of the derived theoretical framework for two paradigmatic physical systems.
Firstly, we show spectral-based moment closure for a prototypical non-Gaussian process: the Bessel process with constant drift, with time dependent parameters \cite{Linetsky2004,linetsky2004spectral}. 
Secondly, we apply our method in the study of the well known phenomenon of stochastic resonance \cite{benzi1981,gammaitoni1998stochastic} where we are able to find some new and non-trivial relations. 
Finally we will discuss the importance of our finding to the data-drive estimation of the Koopman operator in the challenging case of open systems as well as for mean-field models.

\section{Results}

Consider a general dynamical system with a given number of microscopic degrees of freedom.
Instead of following the state dynamics of the whole system, we can equivalently consider the probability density $p(x,t)$ of finding the system in the state $x$ at time $t$.
For a wide class of high-dimensional systems, the dynamics of this probability density is governed by the master equation
\begin{equation*}
    \partial_t p(x,t) = \mathcal{L}_{x} p(x,t) \, .
\end{equation*}

The operator $\mathcal{L}$, generally referred to as the Perron-Frobenius operator \cite{Vulpiani2009-co}, depends on the details of the system.
Here, we consider Fokker-Planck-type equations arising from stochastic diffusive dynamical systems \cite{gardiner2009}.
The approach, however, is fully general and can, in principle, be applied to any master equation in any dimension.
We now work out the expressions for the moments of interest based on the projection of the system state onto the basis determined by the spectrum of the operator $\mathcal{L}_x$ \cite{risken1996}. 
Here, $\lambda_n$ and $\phi_n(x)$ are the eigenvalues and the eigenfunctions of the operator such that
\begin{equation}
\label{eq:Phi}
   \mathcal{L} \ket{\phi_n} = \lambda_n \ket{\phi_n} \, .
\end{equation}
The spectral decomposition consists in writing the time-dependent solution of the original problem as:
\begin{equation}
\label{eq:exp}
    p(x,t) = \phi_0 (x) + \sum_{i=1}^{\infty}a_i(t)\phi_i(x)
\end{equation}
Typically, $\mathcal{L}$ is not Hermitian and for this reason we must also make use of the eigenfunctions $\psi_(x)$ of its adjoint: $\bra{\psi_n} \mathcal{L}^\dagger = \bra{\psi_n} \lambda_n$ \cite{risken1996}. 
Notice that $\braket{\psi_n | \phi_m}=\delta_{mn}$.
Following \cite{mattia2002} one can formally obtain an infinite set of ordinary differential equations (ODE) for the projections $a_n = \langle \psi_n |p\rangle = \int_\mathcal{D} \psi_n(x) p(x,t) dx$, derived directly from the Fokker-Planck equation.
If $\mathcal{L}$ is time independent, then $\phi_0$ represents the stationary probability distribution.
For time-dependent $\mathcal{L}$, we consider $\phi_0$ as the eigenfunction of $\mathcal{L}$ with eigenvalue $\lambda_0 = 0$ assuming that $\mathcal{L}$ depends implicity on time through a set of $K$ parameters $\gamma_i(t)$. 
The system to solve would then be: 
\begin{equation}
\label{eq:ERE}
    \dot{\vec{a}}=\mathbf{\Lambda} \vec{a} + \sum_{i=1}^{K}(\mathbf{C}[{\gamma{_i}]} \vec{a} +\vec{c}[{\gamma_i}]) \dot{\gamma}_i
\end{equation}
where we introduced the diagonal matrix of eigenvalues $\mathbf{\Lambda}$, and the matrix of the coupling integrals $\mathbf{C}[{\gamma}]$ having elements 
\begin{equation}
C[\gamma]_{nm} =- \langle  \psi_n | \partial_{\gamma} \phi_m \rangle 
 = \langle \partial_{\gamma} \psi_n |\phi_m \rangle 
\end{equation}
The same goes for the couplings with the stationary mode $m=0$, $c[\gamma]_n =C[\gamma]_{n0}$, elements of the infinite vector $\vec{c}[\gamma]$.
All this integrals are time-dependent and difficult to estimate even when all the spectral quantities are known, and one must often resort to numerical integration \cite{mattia2002}.

A different approach, introduced in \cite{MattiaVinci2024} and adopted here, considers the dynamical laws governing the evolution of a set of observables $f_n(x)$
\begin{equation}
\label{eq:def_m}
   m_{n}(t)=\braket{f_n|p} = \int_{\mathcal{D}} f_n(x) p(x,t)  dx
\end{equation}
where $\mathcal{D}$ is the support of the probability density $p(x,t)$. 
Applying to \EqRef{eq:def_m} the spectral decomposition (see Appendix for details), a relation between the projections $a_n$ and the observables is obtained:
\begin{equation}
\label{eq:Trick}
    \vec{a}=\mathbf{U^{-1}}(\vec{m} -\vec{m}_*) \, ,
\end{equation}
where the elements of the moment projection matrix $\mathbf{U}$ are
\begin{displaymath}
   U_{nk}=\int_{\mathcal{D}} f_n(x) \phi_k(x)dx
\end{displaymath}
and $\vec{m}_*$ is the vector containing the moments computed with respect to the eigenfunction $\phi_0$:
\begin{displaymath}
   m_{*,n}=\int_{\mathcal{D}} f_n(x) \phi_0(x)dx
\end{displaymath}
We can now close the loop by deriving from the master equation the dynamics of the $n$-th moment: $\dot{m}_n=\braket{f_n|\dot{p}} = \braket{f_n|\mathcal{L}_x p}$.
\begin{displaymath}
   \dot{m}_n = \sum_{k=1}^{\infty}U_{nk} \lambda_k a_k = \mathbf{U} \mathbf{\Lambda} \, \vec{a} ,
\end{displaymath}
which combined with \EqRef{eq:Trick} gives
\begin{equation}
\label{eq:MainResult}
    \dot{\vec{m}} = \mathbf{G_{\lambda}}(t)(\vec{m} -\vec{m}_{*} (t))
\end{equation}
where $\mathbf{G}_\lambda = \mathbf{U} \mathbf{\Lambda}\mathbf{U}^{-1}$.
% \cite{MattiaVinci2024}.

Remarkably, the moment dynamics \eqref{eq:MainResult} holds both for stationary $\mathcal{L}_x$ and for operators changing in time due to their dependence on the system state or on exogenous inputs.
All these conditions are taken into account in the matrix $\mathbf{U}$ and the stationary moments $\vec{m}_*$ through their dependence on the parameters $\vec{\gamma}(t)$.
Moreover, from an analytical point of view, $\mathbf{U}(\vec{\gamma})$ is always constrained by the infinite set of equations (\ref{eq:Phi}), which in some cases can be solved even without knowing the eigenfunctions, as we will show below. 
In what follows, we will mostly choose as observables the moments $f_n(x)=x^n$; however, the specific definition of the variables $\vec{m}$ is irrelevant as long as the matrix $\mathbf{U^{-1}}$ exists.

\section{Eigenvalues of $\mathcal{L}_x$ from the moments dynamics}

The eigenvalues of the operator $\mathcal{L}_{x}$ have a central role in this moment projection framework.
To make this point clear, we consider the Ornstein-Uhlenbeck process whose  Fokker-Planck equation is
\begin{equation}
   \label{eq:OU}
    \partial_t p = \gamma \partial_x(x p) +\frac{\sigma^2}{2} \partial_x^{2}p
\end{equation}
For the first three standard moments ($\langle x^k \rangle$ for $k = 1,2,3$), $\mathbf{G_\lambda}$ can be obtained directly by multiplying (\ref{eq:OU}) by $x^k$ and integrating by parts, yielding:
\begin{displaymath}
    \frac{d}{dt} \begin{pmatrix}
       \langle x \rangle \\
       \langle x^2 \rangle \\
      \langle x^3 \rangle
\end{pmatrix}= \mathbf{G}_\lambda  \begin{pmatrix}
       \langle x \rangle \\
       \langle x^2 \rangle -\sigma^2\\
      \langle x^3 \rangle
\end{pmatrix}
\end{displaymath}
where $\langle x^n \rangle = \braket{x^n|p}$ and we defined
\begin{displaymath}
    \mathbf{G}_\lambda = \begin{pmatrix}
       -\gamma & 0  & 0 \\
       0 & -2\gamma &0 \\
       3\sigma^2 & 0 & -3\gamma
\end{pmatrix}
\end{displaymath}
whose eigenvalues are $\lambda_n=-n\gamma$, which coincides with the eigenvalues of the Fokker-Planck operator \cite{risken1996}. 
Consequently the relaxation time scales of the moments dynamics coincides with the eigenvalues of $\mathcal{L}_x$.
A more interesting system is the multidimensional case:
$$\partial_t p(\vec{x},t)=\sum_{i,j}\left [ \partial_{x_i}M_{ij}x_j +\sum_j D_{ij}\partial_{x_i}\partial_{x_j} \right] p(\vec{x},t)$$
Using the set of moments $m_k=\braket{x_k}$ for $k=1,..,n$, we find that $\mathbf{G}_{\lambda}=-\mathbf{M}$; hence, the first $n$ eigenvalues of the operator
$\mathcal{L}_x$ are the eigenvalues of the drift matrix $\mathbf{M}$.
In the particular case of the Langevin equation in a parabolic potential:
$$\partial_t p(x,v,t)=-\partial_x(vp) + \partial_v{(\gamma v + \omega^2 x)p} +D \partial_v ^2 p$$
a convenient set of observables is ($\langle x \rangle$, $\langle v\rangle$, $\langle xv \rangle$, $\langle v^2 \rangle$, $\langle x^2 \rangle$), which leads to the following $\mathbf{G}_\lambda$:
\begin{displaymath}
	\begin{pmatrix}
0 & 1  & 0 & 0 &0 \\
-\omega^2& -\gamma  & 0 & 0 & 0\\
0 & 0  & -\gamma  &  1& -\omega^2\\
0 & 0  &-2\omega^2  & -2\gamma  & 0 \\
0 & 0  &2  & 0 & 0
  \end{pmatrix} \, .
\end{displaymath}
Here we have included only these five moments because
this specific choice of observables leads to a block matrix which in turn implies moment closure. 
The first block with eigenvalues  $\lambda_{1,2}=\frac{-\gamma \pm \sqrt{\gamma^2 - 4\omega^2}}{2}$, while the eigenvalues of the second block matrix satisfy: $\lambda  \left(2 \gamma ^2+3 \omega ^2\right)+3 \gamma  \lambda ^2+2 \gamma  \omega ^2+\lambda ^3 =0$, whose solutions are simply $\lambda = \lambda_1 +\lambda_2 , 2\lambda_1 , 2\lambda_2$ 

This way of obtaining the eigenvalues of $\mathcal{L}$ is appealing because no knowledge of the eigenfunctions is needed.

\section{Data-driven estimation of $\mathbf{G_{\lambda}}$ and the Koopman operator}

Most of the time, when working with complex systems, the dynamical laws governing the single agents are not known.
Nowadays, however, we have access to large amounts of data, which has opened the doors to data-driven approaches that aim to reconstruct emergent laws directly from experimental data.
In this context, a promising framework is the formalism first introduced by Koopman long ago \cite{koopman1931hamiltonian,koopman1932dynamical}, which has found renewed interest in recent years \cite{mezic2005spectral}.
The idea applies to both generic nonlinear deterministic dynamical systems and stochastic systems like the one we study in this paper.
In this brief section, we aim to highlight the deep connections between this approach and the results we have presented so far.
The Koopman operator, $\mathcal{K}$, is an infinite-dimensional linear operator that governs the evolution in time of observables of a stochastic system.
So, following the Extended Dynamic Mode Decomposition (EDMD) introduced in \cite{williams2015data,di2024non}, we first choose a dictionary of $N$ functions $\{f_k(\mathbf{x})\}_{k=1}^{N}$.
The observables are defined as expectation values of the functions $f_k(\mathbf{x})$ with respect to the time-dependent measure of the system, or, in other words, what we have so far called moments: $m_k(t)=\int_{\mathcal{D}}f_k(\mathbf{x})p(\mathbf{x},t)d\mathbf{x}$.
Formally speaking, if we take $N \to \infty $, then the single time-step evolution of the vector of observables is given by:
$$\vec{m}(t + dt)=\mathcal{K}\vec{m}(t)$$
Moreover, if $p(\mathbf{x},t)$ satisfies the differential equation:
$$\partial_t p = \mathcal{L} p$$
then $\mathcal{K}=e^{\mathcal{L}^\dagger dt}$, where $^\dagger$ indicates the adjoint operation \cite{mezic2005spectral}.
Specifically, if the equation for $p(\mathbf{x},t)$ is of the Fokker-Planck type, then $\mathcal{K}$ is the operator of the backward Kolmogorov equation \cite{risken1996}, which is typically used in the study of first-passage times \cite{bray2013persistence}.
Now consider the system \eqref{eq:MainResult} in the autonomous case. 
It is convenient to include in the dictionary $\vec{m}$ of moments also the constant observable $m_1 = \int p(x,t) dx=1$. 
In this way the first column  as $\mathbf{G}_{1,1}=0$ and $\mathbf{G}_{1,n}= - m_{*,n}$ which allows us to rewrite \EqRef{eq:MainResult} as:
\begin{equation}
\label{eq:new6}
    \dot{\vec{m}}=\mathbf{G_{\lambda}}\vec{m} 
\end{equation}
that has an analytical solution:
$$\vec{m}(t+dt)=\mathbf{e^{G_{\lambda}dt} }\vec{m}(t)$$
Thus, $\mathcal{K}=\mathbf{e^{G_{\lambda}dt}}$. Moreover, since $\mathcal{K}=\mathbf{U}\mathbf{e^{\Lambda dt}}\mathbf{U^{-1}}$, the left eigenvectors of $\mathcal{K}$ are the columns of $\mathbf{U}$. The data-driven estimation of the Koopman operator is an active field of research; however, in most cases, the system studied is autonomous. 
We have already pointed out, however, that once the parametric dependence of $\mathbf{G}_{\lambda}$ on the parameters $\gamma$ is established, we can then also investigate the non-autonomous scenario where $\gamma(t)$ as a function of time.

\section{Open systems}

A crucial aspect to consider is understanding what may go wrong when truncating the system of moments in the presence of external forcing, in order to better appreciate the limitations of this approach. 
Consider equation \EqRef{eq:ERE} for a single time-dependent parameter $\gamma(t)$. 
For simplicity, we include the mode $a_0(t)=1$ (corresponding to the eigenfunction $\phi_0$) in the infinite vector $\vec{a} =(a_0,a_1,\dots)$, yielding: 
\begin{equation}
\label{eq:ERE2} 
   \dot{\vec{a}}=\mathbf{\Lambda}\vec{a} +\dot{\gamma}\mathbf{C}\vec{a} \, . 
\end{equation} 
Unlike the autonomous case ($\dot{\gamma}=0$), in this non-autonomous scenario, the dynamics of a given mode $a_n$ generally depends on all the others due to the coupling matrix $\mathbf{C}$, which mixes the modes in \EqRef{eq:ERE2}). 
Therefore, truncating the spectral decomposition to the first $n$ modes introduces an error proportional to the magnitude of the neglected terms in $\mathbf{C}$.

In the following we  quantify the error by comparing the the truncated moment system against the truncated spectral expansion.
Consider a set of functions $f_{k}(x)$ and their expectation values defined as $m_k(t)= \braket{f_k|p}$. 
For notational simplicity, also in this case we include the moment $\langle p \rangle=1$, and we define the infinite vector $\vec{m}=(1,m_1,\dots)$. 
Taking the time derivative of $\vec{m}=\mathbf{U}\vec{a}$ we get:
$$\dot{\vec{m}}= \dot{\gamma} \partial_{\gamma}\mathbf{U} \vec{a} +\mathbf{U}\dot{\vec{a}}$$
using eq. (\ref{eq:ERE2}) and the fact that $\vec{a}=\mathbf{U}^{-1}\vec{m}$ we arrive at the expression:
\begin{equation}
\label{eq:expandedM}
   \dot{\vec{m}} = \mathbf{G}\vec{m} + \dot{\gamma}\mathbf{U}(-\partial_{\gamma} \mathbf{U^{-1}} +\mathbf{C}\mathbf{U^{-1}})\vec{m} \, .
\end{equation} 

This equation indicates that when $\dot{\gamma}\neq 0$, (\ref{eq:new6}) and (\ref{eq:expandedM}) are equivalent only if: 
\begin{equation}
\label{eq:CvsU} 
   \mathbf{C}=\partial_{\gamma} \mathbf{U^{-1}}\mathbf{U} 
\end{equation}
We remark that \EqRef{eq:CvsU} tell us something non trivial: while  $\mathbf{C}$ is entirely determined by the eigenfunctions of the operators $\mathcal{L}$ and $\mathcal{L^{\dagger}}$,  the matrix $\mathbf{U}$ depends heavily on the specific choice of observables $f_k(x)$. 
Hence, \EqRef{eq:CvsU} can be used, as we will show below, as an indicator of the goodness of the choice of the dictionary $f_k(x)$ which is central for all data-driven approaches \cite{klus2020data}. 

Let us consider for a moment an autonomous system. 
In this case, we can express the solution  discussed above using the exponential matrix. 
In particular, we have:
\begin{equation}
\label{eq:DynamicsUncoupled}
\begin{split}
\vec{a}(t )&=e^{\mathbf{\Lambda} t}\vec{a}(0) \\
\vec{m}(t) &= \mathbf{U} e^{\mathbf{\Lambda} t}\mathbf{U}^{-1} \vec{m}(0)\\
\end{split}
\end{equation}
These expressions are valid even if we truncate the spectral expansion to a finite term $N <\infty$. Then, from (\ref{eq:DynamicsUncoupled}), we can understand under what condition the truncated moment system is equivalent to the truncated eigenmode system. In fact, it is sufficient to require that at the initial time $t=0$, $\mathbf{U}^{-1} \vec{m}(0)=\vec{a}(0)$. In fact, in this case:
$$ \mathbf{U}^{-1}\vec{m}(t)= e^{\mathbf{\Lambda} t}\mathbf{U}^{-1} \vec{m}(0)=e^{\mathbf{\Lambda} t}\vec{a}(0)=\vec{a}(t)$$
The important remark  is that is not always possible to satisfy this equivalence for arbitrary initial conditions if we truncate the moment systems.

To better clarify this point, consider the initial condition $p(x,t=0)=\delta(x-x')$. In this case, the initial moments are $m_k(0)=f_k(x')$ while the eigenmodes are $a_k(0)=\psi_k(x')$. 
In other words, $\mathbf{U}^{-1} \vec{m}(0)=\vec{a}(0)$ implies:
\begin{equation}\label{eq:PsiReconstruction}
    \psi_i(x')=\sum_j^N U_{ij}^{-1}f_j(x')
\end{equation}
From (\ref{eq:PsiReconstruction}), we see that exact equivalence of initial conditions for the truncated systems is possible only if it is possible to express the $N$ adjoint eigenfunctions as a finite series of the chosen basis library $\{f_k\}_{k=1,..,N}$. Moreover, if we now take the derivative with respect to $\gamma$ and then integrate both sides of (\ref{eq:PsiReconstruction}) with $\int dx' \phi_k(x')$, we recover equation (\ref{eq:CvsU}). We can then conclude that the moment system and spectral expansion are in this case equivalent even in the non-autonomous case. For example let us  consider a case where all the integrals can be carried out analytically, i.e., the Ornstein–Uhlenbeck process (\ref{eq:OU}). The adjoint eigenfunctions are the Hermite polynomials \cite{risken1996} $\psi_n(x)=H_n(x \frac{\sqrt{\gamma}}{\sigma})$, which are a linear combination of monomials $f_k(x)=x^k$; hence, the standard moments satisfy (\ref{eq:PsiReconstruction}). The projection matrix for the first 5 moments, including $k=0$, and its inverse are:
\begin{equation} \label{eq:UOU}
\mathbf{U}=\left(
\begin{array}{ccccc}
 1 & 0 & 0 & 0 & 0 \\
 0 & \frac{\sigma }{2 \sqrt{\gamma }} & 0 & 0 & 0 \\
 \frac{\sigma ^2}{2 \gamma } & 0 & \frac{\sigma ^2}{4 \gamma } & 0 & 0 \\
 0 & \frac{3 \sigma ^3}{4 \gamma ^{3/2}} & 0 & \frac{\sigma ^3}{8 \gamma ^{3/2}} & 0 \\
 \frac{3 \sigma ^4}{4 \gamma ^2} & 0 & \frac{3 \sigma ^4}{4 \gamma ^2} & 0 & \frac{\sigma ^4}{16 \gamma ^2} \\
\end{array}
\right)
\end{equation}
\begin{equation*}
\mathbf{U^{-1}}=  \left(
\begin{array}{ccccc}
 1 & 0 & 0 & 0 & 0 \\
 0 & \frac{2 \sqrt{\gamma }}{\sigma } & 0 & 0 & 0 \\
 -2 & 0 & \frac{4 \gamma }{\sigma ^2} & 0 & 0 \\
 0 & -\frac{12 \sqrt{\gamma }}{\sigma } & 0 & \frac{8 \gamma ^{3/2}}{\sigma ^3} & 0 \\
 12 & 0 & -\frac{48 \gamma }{\sigma ^2} & 0 & \frac{16 \gamma ^2}{\sigma ^4} \\
\end{array}
\right)
\end{equation*}
which, using (9), leads to:
\begin{equation*}
   \partial_{\gamma}\mathbf{U}^{-1}\mathbf{U}= \left(
\begin{array}{ccccc}
 0 & 0 & 0 & 0 & 0  \\
 0 & \frac{1}{2 \gamma } & 0 & 0 & 0  \\
 \frac{2}{\gamma } & 0 & \frac{1}{\gamma } & 0 & 0  \\
 0 & \frac{6}{\gamma } & 0 & \frac{3}{2 \gamma } & 0  \\
 0 & 0 & \frac{12}{\gamma } & 0 & \frac{2}{\gamma } \\
 \end{array}
\right)=\mathbf{C}
\end{equation*}
in accordance with what we would get using the definition of $\mathbf{C}$. 
Note that the matrix $\mathbf{U}$ has a peculiar property which reflects the fact that this choice of observables solves the moment-closure problem: the inverse of the truncated $\mathbf{U}$ is equivalent to the truncated inverse. 
If we instead choose the library $f_k^e (x)=e^x x^k$, we do not have moment closure. 
In fact, using (\ref{eq:OU}), it can be shown that $\dot{m_k}$ depends on itself and on $m_j$ with $j=k-1,k-2,k+1$. 
If we compute the truncated approximation of $\mathbf{C}$, we have instead:
\begin{equation*}
    \partial_{\gamma}\mathbf{U}^{-1}\mathbf{U}=\left(
\begin{array}{ccccc}
 0 & 0 & 0 & \frac{\sigma ^5}{192 \gamma ^{7/2}} & \frac{5 \sigma ^6}{1536 \gamma ^4} \\
 0 & \frac{1}{2 \gamma } & 0 & -\frac{5 \sigma ^4}{96 \gamma ^3} & -\frac{\sigma ^5}{32 \gamma ^{7/2}} \\
 \frac{2}{\gamma } & 0 & \frac{1}{\gamma } & \frac{5 \sigma ^3}{12 \gamma ^{5/2}} & \frac{15 \sigma ^4}{64 \gamma ^3} \\
 0 & \frac{6}{\gamma } & 0 & \frac{3 \gamma -5 \sigma ^2}{2 \gamma ^2} & -\frac{5 \sigma ^3}{4 \gamma ^{5/2}} \\
 0 & 0 & \frac{12}{\gamma } & \frac{10 \sigma }{\gamma ^{3/2}} & \frac{8 \gamma +15 \sigma ^2}{4 \gamma ^2} \\
\end{array} \right)
\end{equation*}
which shows how an error in \EqRef{eq:PsiReconstruction} inevitably leads to an error in the estimated coupling matrix $\mathbf{C}$. 
Note that, by construction, these two sets of observables yield a matrix $\mathbf{G}$ with the same eigenvalues $\lambda$. 
This is an important point: studying the relaxation dynamics of the autonomous case, as it is typically done within the Koopman framework \cite{klus2020data}, it might lead to the impression that two different choices of basis functions with same spectrum of eigenvalues, as the simple example above, are both reasonably good because the relaxation time scales are indeed the same. 
However, we showed that it is crucial to investigate and validate how a specific choice of observable behaves under time dependent modulation. 
For this reason we believe that \EqRef{eq:expandedM} will be particularly relevant in the data-driven estimation of the Koopman operator.

\begin{figure}
    \centering
    \includegraphics{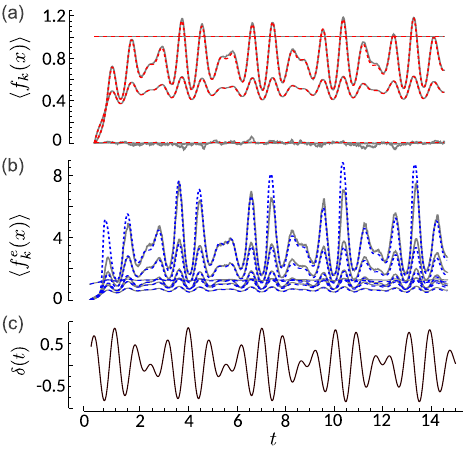} % Assuming image file name is MOMENTSDYNAMICS.pdf or similar
    \caption{(a) shows the time-dependent dynamics of the first five moments $m_k=\langle x^k\rangle$ with $k=0,..,4$ of the Ornstein–Uhlenbeck process described by \EqRef{eq:OU} with a time-dependent $\gamma(t)=1+\delta(t)$ as shown in (c) and $\sigma=1$. 
   The truncated moment system (red) is in perfect agreement with Monte Carlo simulations (grey). 
   The same is not true if we use as moments $m_k=\langle x^k e^x\rangle$, as shown in figure (b) (blue). }
\label{fig:OU}
\end{figure}

However, as shown in \FigRef{fig:OU}, when the parameter $\gamma(t)$ varies in time, the higher-order exponential moments $\langle f_k^e(x) \rangle$ are not accurately captured by the truncated moment system, in contrast to the standard moments $\langle f_k(x) \rangle$.
In a practical scenario, this could mean that two sets of dictionaries may yield comparable errors in the autonomous case--that is, the eigenvalue estimates are similarly accurate when using, for example, the EDMD approach--yet one of them may perform substantially worse once the system is coupled to an external environment.
We anticipate that the results discussed in this section will be useful for identifying such limitations and for improving data-driven strategies for the control of stochastic systems, which represent one of the most promising applications of the Koopman approach \cite{kaiser2021data}.

\section{Moments closure through spectral expansion}

\begin{figure*}
    \centering
   \includegraphics{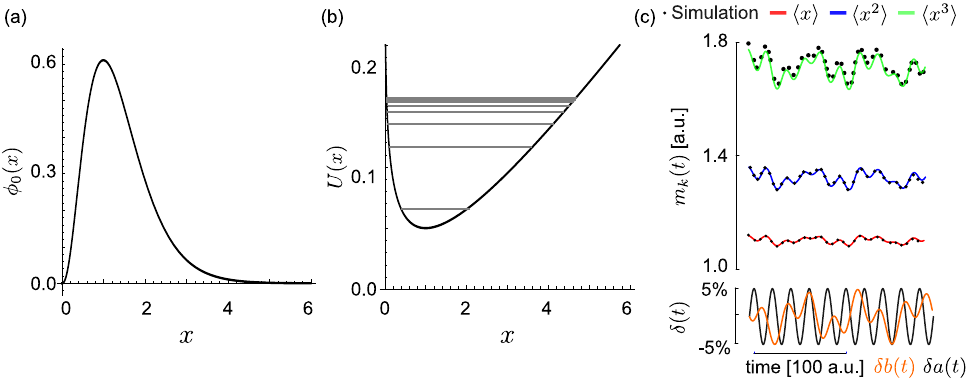}
    \caption{Stationary solution, (a), and potential, (b), of the Bessel process with constant drift described by equation (\ref{eq:BesselFP}).
    The dominant eigenvalues of the Fokker-Planck operator are also shown, with a proper shift and rescaling, in figure (b) (gray).
    The parameters are: $a=-0.05,b=0.05$ and $\sigma=0.2$.
    In figure (c), we compare the results of a detailed Monte Carlo simulation (black dots) and a 3D moment closure theory discussed in the text (colored trajectories for $\langle x^k \rangle_{t}$).
    In this case, the system is non-autonomous since both parameters $a$ and $b$ are modulated in time, as shown in the bottom panel.
    In particular, $a(t) =a_0 (1 +\delta a(t))$ and $b(t) =b_0 (1 +\delta b(t))$, where $a_0 =-0.05$, $b_0=0.05$ and $\sigma =0.1$, while $\delta a (t) =\sin(2\pi t\epsilon_1)\epsilon_2/2$, $\delta b (t) =[\sin(0.783212\pi t\epsilon_2) +\sin(0.31672 \pi t\epsilon_2)]\epsilon_1/4$, with $\epsilon_1=0.1$ and $\epsilon_2=0.05$.
    }
    \label{fig:besselpdf}
\end{figure*}

We now return to the original problem of closing the moments system. When the operator $\mathcal{L}$ is nonlinear, the hierarchy of equations cannot generally be closed on a finite number of moments. This raises the question as to what is the best procedure to close and thus approximate the equations, and is typically referred to as the moment closure problem. A common approximation made, for example, is a Gaussian Ansatz for $p(x,t)$. This automatically ensures that all the central moments of order $n>2$ are zero and leads to self-consistency relations that allow expressing the unknown moments in terms of the known ones. Another possibility is to assume some kind of separation of time scales such that the higher moments are much faster than the lower ones, and with an adiabatic approximation, the system is closed. Both these approaches and their more elaborate variants have several issues. 

The most prominent is the impossibility of justifying the main assumption in the general scenario such that, even if many cases of success have been reported, very few of them could be explained and generalized \cite{kuehn2016}. The results of this paper give a proper framework for investigating the moment-closure problem. Equation (\ref{eq:MainResult}) is general and applies even to open, out-of-equilibrium systems thanks to the moving basis of $\mathcal{L}$. To show it in action, we pick a prototypical system where standard Gaussian closure fails: the Bessel process with constant drift, which has many applications in physics \cite{toppaladoddi2015theory,toppaladoddi2017statistical}, finance \cite{linetsky2004spectral}, and biology \cite{fogedby2007dynamics}. The Fokker-Planck equation describing the process is:
\begin{equation}\label{eq:BesselFP}
    \partial_t p(x,t) =-\partial_x \left [ \left ( \frac{b}{x} +a \right ) p(x,t) \right ] + \frac{\sigma^2}{2} \partial_x^2 p(x,t)
\end{equation}
defined for $x>0, a<0$, and with a reflective barrier at the origin.
Figure \ref{fig:besselpdf} shows the stationary solution and eigenvalue spectrum for a specific choice of parameters.
From (\ref{eq:BesselFP}), we can get the equation for the mean $\langle x \rangle$:
$$ \dot{\langle x \rangle}= b \left \langle \frac{1}{x} \right \rangle + a $$
So in this case, the moment $\langle x^n \rangle$ will depend on the moments $\langle x^{n-1} \rangle$ and $\langle x^{n-2} \rangle$, making it impossible to use Gaussian closure.
On the other hand, for this system, the matrix $\mathbf{G_{\lambda}^{(k)}}$ has an analytical expression up to any order $k$ of truncation of the expansion, as we will show. Note that for this process, the eigenvalues and eigenfunctions of the operator have been found \cite{guarnieri2017}; however, this does not guarantee the analytical closure of the standard spectral expansion (\ref{eq:ERE}) in the non-autonomous case. To show this, consider the case in which an external drive is changing the parameter $b(t)$ in time in (\ref{eq:BesselFP}). Then, to close the system (\ref{eq:ERE}), we would have to compute $\partial_b \psi_{n}(x,a,b)$, which in this specific case would require knowing $\partial_{k} L_n^k(x)$. Unfortunately, no closed formula for the derivative with respect to the parameters of the Laguerre function, $L_n^k(x)$, is known, and we must then resort to numerical computation of $k^2 +k$ integrals for each value of $b(t)$, making the standard approach unfeasible. If we consider instead the moment projection system (\ref{eq:MainResult}), we can calculate everything analytically using the eigenfunction definition (\ref{eq:Phi}). For the null mode, $\lambda=0$, this leads to a closed recurrence relation:

\begin{equation*}
a m_{*,k-1} + \left[ \frac{\sigma^2}{2}(k-1) +b \right]m_{*,k-2}=0
\end{equation*}
with the normalization condition $m_{*,0}=1$ while $m_{*,1}=-\frac{b + \sigma^2 /2}{a}$. For the non-stationary projections, we have instead:
\begin{equation*}
    U_{k,n}=\frac{k}{\lambda_n}\left[ a U_{k-1,n}  +\frac{\sigma^2  (k-1) +2b}{2} U_{k-2,n} \right]
\end{equation*}
since $ U_{0,n}=0$, we just need to compute one integral to close the system, $ U_{-1,n}$. However, it turns out, quite surprisingly, that this step is not really needed. In fact, the unknown $U_{-1,n}$ cancels out when computing $\mathbf{G_{\lambda}}=\mathbf{U} \mathbf{\Lambda} \mathbf{U^{-1}}$ at any order of the truncation. For example, in the two-dimensional closure $\vec{m}= (\langle x \rangle,\langle x^2 \rangle)$, we have:
\begin{equation*}
\mathbf{U}=\begin{pmatrix}
U_{11} & U_{12} \\
\frac{2 a U_{11}}{\lambda_1} & \frac{2 a U_{12}}{\lambda_2}
\end{pmatrix}\;\;\;
\mathbf{G}=\begin{pmatrix}
\lambda_1 +\lambda_2 & -\frac{\lambda_1\lambda_2}{2a} \\
 2a & 0
\end{pmatrix}
\end{equation*}
while a three-dimensional closure $\vec{m}= (\langle x \rangle,\langle x^2 \rangle,\langle x^3 \rangle)$ leads to:
\begin{equation}\label{eq:MomentClosureBessel}
\mathbf{G}=\begin{pmatrix}
\lambda_1 +\lambda_2 + \lambda_3 & c &\frac{\lambda_1\lambda_2\lambda_3}{6 a^2}\\
 2a & 0 & 0 \\
3(b +\sigma^2 ) & 3 a & 0
\end{pmatrix}
\end{equation}
with $c=-\frac{2 a^2(\lambda_2\lambda_3 +\lambda_1(\lambda_2 + \lambda_3)) +(b +\sigma^2)\lambda_1\lambda_2\lambda_3}{4 a ^3}$. Note that the last two rows of $\mathbf{G}$ can also be obtained directly from the Fokker-Planck equation. The eigenvalues of the Fokker-Planck operator in (\ref{eq:BesselFP}) are (see \cite{guarnieri2017} for the derivation): $\lambda_n =-\frac{2a^2 n b +a^2 n^2 \sigma^2}{2(b+n\sigma^2)^2}$; hence, the moment system is fully closed at any order, i.e. number of modes included. So far, we have dealt with autonomous systems, but the great advantage of the framework we present here is that it is much simpler with respect to standard spectral expansion when the parameters of the continuity equation vary with time. In fact, once the parametric dependence of $\mathbf{G_\lambda}$ on the time-varied parameters is established, we can use it to describe open systems as well. However, a general remark is required. Truncation of the spectral expansion to the order $n$ means neglecting the time scales of the system that are faster than $1/\mathrm{Re}(\lambda_n)$. As we have shown here, moment closure systems and spectral expansion are indeed two faces of the same coin. Clearly then, a truncated system can describe well a non-autonomous case only if the external time-dependent perturbation is slower than the last time scale included. In figure \ref{fig:besselpdf}, we show how the system of three moments from eq. (\ref{eq:MomentClosureBessel}) successfully reproduces the time-dependent solution, obtained with Monte Carlo simulations, for a case in which both parameters, $a$ and $b$, are differently modulated in time. In other words, with the sole knowledge of the eigenvalue spectrum of the unperturbed system, we have solved analytically the moment closure problem for the time-dependent open system. Of course, the eigenvalues $\lambda_n$ carry information about the eigenfunctions; however, they can also be derived directly from the relaxation to equilibrium of the unperturbed system. We will come back to data-driven estimation of $\mathbf{G}$ in the following sections. \\
While it is difficult to find, for the most general case, a choice of observables that greatly simplifies $\mathbf{G}$.If we keep only the first mode in the expansion, $G_{\lambda}=\lambda_1$, independently of the system under study. This is a crude approximation, valid only very close to the stationary state. If we reduce the generality and consider only processes governed by Fokker-Planck equations, we can find a general two-dimensional moment system. Take the continuity equation for a $n$-dimensional process $\vec{x}$:
\begin{equation}\label{eq:GeneralFP}
    \partial_t p(\vec{x})=-\sum_{i}^{n}\partial_{x_i}f_i(\vec{x})p(\vec{x}) + \sum_{i,j}^{n}\partial_{x_i}\partial_{x_j}D_{ij}(\vec{x})p(\vec{x})
\end{equation}
with the diffusion matrix $D_{ij}(\vec{x})$ positive definite in $\mathcal{D}$. If we choose the moments $m_i(t)=\int_{\mathcal{D}} \prod_{j}^{n}dx_j x_i p(\vec{x},t)$ and $y_i(t)=\int_{\mathcal{D}} \prod_{j}^{n}dx_j f_i(\vec{x}) p(\vec{x},t)$, the matrix $\mathbf{G}$ simplifies greatly. To see this, using the equation for the eigenfunction $\phi_n (\vec{x})$ and integrating by parts, it can be shown that:
$$\lambda_n \int_{\mathcal{D}} \prod_{j}^{n}dx_j x_i \phi_n(\vec{x},t)=\lambda_n U_{in}=U_{jn} + b.t.$$
where $U_{2n}=\int_{\mathcal{D}} \prod_{j}^{n}dx_j f_i \phi_n(\vec{x},t)$ and $U_{1n}=\int_{\mathcal{D}} \prod_{j}^{n}dx_j x_i \phi_n(\vec{x},t)$. Under natural boundary conditions, the boundary term $b.t.$ vanishes; hence, the matrices read:

\begin{equation}\label{eq:G2m}
\mathbf{U}=\begin{pmatrix}
U_{11} & U_{12} \\
\frac{U_{11}}{\lambda_1} & \frac{ U_{12}}{\lambda_2}
\end{pmatrix}\;\;\;
\mathbf{G}=\begin{pmatrix}
0 & 1 \\
 -\lambda_1 \lambda_2 & \lambda_1 +\lambda_2
\end{pmatrix}
\end{equation}

As in the case of the Bessel process, the unknown terms $U_{11},U_{12}$ cancel out when computing \textbf{G}, which is surprising given the general premises of equation (\ref{eq:GeneralFP}). For this two moments we thus have the system of equations:
\begin{equation}
    \begin{split}
        & \dot{m}_i= (y_i-y_{i,*}) \\
        & \dot{y}_i=-\lambda_1 \lambda_2 (m_i-m_{i,*}) + (\lambda_1 +\lambda_2) (y_i-y_{i,*}) \end{split}
\end{equation}

\section{Stochastic Resonance}
\begin{figure}[ht]
    \centering
   \includegraphics{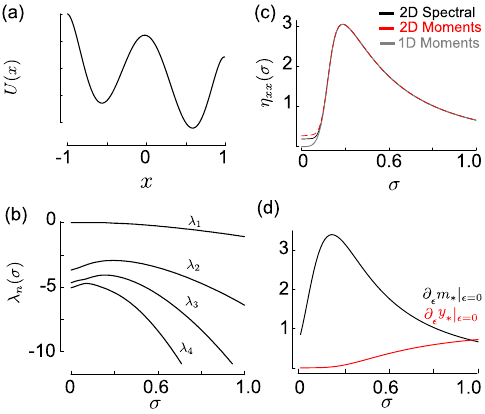}
    \caption{Figure (a) shows the bistable potential used in the text $U(x)=-4(x^2 /2 -x^4 +x^6 /2 +\beta x )$ for $\beta=0.02$. For this potential, we report in figure (b) the first 4 nonzero eigenvalues as a function of the standard deviation of the noise term $\sigma$. Note the quasi-degeneracy of $\lambda_1$ with respect to $\lambda_0=0$ for small noise intensity, typical of multistable potentials. 
    In Figure (c), the spectral gain for modulation frequency $\Omega =0.01$ is shown as a function of the noise intensity using the different approaches discussed in the text, i.e., the 1D and 2D Moment systems (gray and red) and the 2D spectral expansion (black). Figure (d) shows all the parameters needed for the computation of the spectral gain as a function of the noise standard deviation $\sigma$.}
    \label{fig:BistableEig}
\end{figure}

We can use the results of previous sections to study stochastic resonance in a rather general context. To do so, consider a bistable system under weak periodic forcing $\epsilon(t)= \epsilon_1 sin(\Omega t)$ with $\epsilon_1 \ll 1$ \cite{benzi1981}. The continuity equation is of the Fokker-Planck type:
\begin{equation*}
\mathbf{L_x} \cdot= \partial_x[(\partial_x U(x) + a(x)\epsilon(t) )\cdot] +\frac{\sigma^2}{2} \partial_{x}^{2}\cdot
\end{equation*}
As a benchmark, we consider the potential $U(x)$ used in \cite{williams2015data} and reflective boundary conditions at $x =\pm 1$, see Fig. \ref{fig:BistableEig} (a).
 This kind of system can exhibit the phenomenon of stochastic resonance: a specific value of noise intensity $\sigma$ exists, such that the frequency of noise-induced hopping between the two different states is maximized. This peculiar resonance can be studied by examining the linear response of the moments. In particular, if the periodic stimulation is weak, we can make the Ansatz $\vec{m}(t)=\vec{m}_* +\epsilon_1 \vec{m}_1(t)$; thus, $\vec{m}_1$ is the fluctuation with respect to the stationary state. The amplification factor for each moment considered is then defined as the absolute value, in the frequency domain, $|\frac{m_1 (\omega)}{\epsilon(\omega)}|$. Applying linear response theory to (\ref{eq:MainResult}), we get:
 \begin{equation}\label{eq:AplificationFactor}
     \vec{\eta}(\Omega)=\left | (\mathbf{I}i\Omega -\mathbf{G^{0}})^{-1}\mathbf{G^{0}}\partial_{\epsilon}\vec{m}_*|_{\epsilon=0}  \right |
 \end{equation}
here the absolute value is element-wise and the suffix $^0$ indicates quantities related to the unperturbed system ($\epsilon_1=0$).
If the angular frequency of the stimulation is such that $\Omega < |\lambda_1|$, a reliable approximation is to consider a single mode in the expansion. In this case, $\mathbf{G_\lambda}=\lambda_1$ for any choice of observable, $m=\langle g(x)\rangle$, with an arbitrary $g(x)$. Moreover, for one-dimensional systems, and more generally when the drift term can be expressed as a gradient of a potential, the stationary solution takes the Boltzmann form $\phi_0 (x) \sim e^{-2(U(x) +\epsilon A(x))/\sigma^2}$, with $a(x)=\partial_x A(x)$. In this case, it can be shown that:
\begin{equation*}
\begin{split}
\partial_\epsilon \langle g(x,\epsilon) \rangle_0|_{\epsilon=0} &= \frac{2}{\sigma^2}[\langle A(x)g(x) \rangle_0 -\langle A(x) \rangle_0 \langle g(x) \rangle_0]_{\epsilon=0}\\
& \stackrel{\text{def}}{=}\frac{2}{\sigma^2}\langle \langle g(x)A(x) \rangle \rangle_0
\end{split}
\end{equation*}
In other terms, the response of the perturbed system is tightly related to fluctuations of the unperturbed one, an argument typically referred to as \textit{fluctuation theorem} \cite{gammaitoni1998stochastic}.
Putting everything together (see Appendix), we arrive at a very general formula for the amplification factor of an observable $g(x)$ under the stimulation with amplitude $a(x)=\partial_x A(x)$:
\begin{equation}\label{eq:SR}
    \eta_{g,A}(\Omega,D)=\left | \frac{\lambda_1 (\sigma)}{\lambda_1 (\sigma) - i\Omega}\frac{2\langle \langle g(x)A(x) \rangle \rangle_0 }{\sigma^2} \right |
\end{equation}
both the eigenvalue and the expectation value with respect to the stationary state are for the unperturbed system ($\epsilon_1 =0$). This result is interesting for different reasons. Equation (\ref{eq:SR}) is symmetric for the exchange $\eta_{g,A}=\eta_{A,g}$, a property that, to the best of our knowledge, was never noted before. Secondly, in the literature, it is common to find an expression for the amplification of the mean under constant amplitude, i.e., $\eta_{x,x}$, which is a specific case of the previous results:
\begin{equation}\label{eq:MeanResponse}
    \eta_{x,x}=\left | \frac{\lambda_1(\sigma)}{\lambda_1 -i\Omega}\frac{2\mathrm{Var}(x)}{\sigma^2} \right|
\end{equation}
with $\mathrm{Var}(x)$ being the variance of $x$. Proving the same formula in the standard spectral expansion framework, i.e., eq. (\ref{eq:ERE}), is possible only perturbatively and requires laborious calculation to obtain an approximation of the eigenfunctions of $\mathcal{L}$ and its adjoint \cite{gang1990periodically,fox1989}.
Separation of time scales in the system is encoded in the spectral gap, i.e., the distance between successive eigenvalues, and it is clearly what justifies a truncation of the series we have used so far. In the context of stochastic resonance, however, the frequency of the periodic forcing is typically very slow. At the same time, for small noise intensity $\sigma$, the dominant eigenvalue is generally almost degenerate, i.e., $\lambda_1 \approx \lambda_0 =0$, see Fig. \ref{fig:BistableEig} (b), which would imply $|\lambda_1| < \Omega$. Computing the correction to the previous result to include $\lambda_2$ requires a lot of effort using the spectral expansion for the reasons we have already discussed. On the contrary, using the moments approach, it is straightforward thanks to the result (\ref{eq:G2m}). In fact, choosing as the second moment $y=\int_{\mathcal{D}}(\partial_x U(x) +\epsilon(t)a(x)) dx$ leads to:
\begin{equation}\label{eq:Eren}
    \begin{split}
        & \dot{m}= (y-y_*) \\
        & \dot{y}=-\lambda_1 \lambda_2 (m-m_*) + (\lambda_1 +\lambda_2) (y-y_*) -\dot{\epsilon}(t)\braket{a(x)}
    \end{split}
\end{equation}
which is by itself a novel and interesting result reminiscent of the Ehrenfest theorem in quantum mechanics. Using (\ref{eq:Eren}), we can extract the linear response and compute the correction to eq. (\ref{eq:MeanResponse}):
\begin{equation}\label{eq:2lambResponse}
    \eta_{x,x}= \left | \frac{i\Omega(\partial_{\epsilon}y_*|_{\epsilon=0}) - \lambda_1 \lambda_2\partial_{\epsilon}m_*|_{\epsilon=0}}{(i\Omega-\lambda_1) (i\Omega-\lambda_2)} \right |
\end{equation}
Note that $\partial_{\epsilon}y_*|_{\epsilon=0}=\braket{a(x)} +\frac{2}{\sigma^2} \langle\langle \partial_x U(x) a(x) \rangle \rangle_0$, see Fig. \ref{fig:BistableEig} (d) for its value as a function of $\sigma$. In Fig. \ref{fig:BistableEig} (c), we compare all the different expressions that we have reported here for $\eta_{x,x}$ to the exact amplification computed from the time-dependent solution of the Fokker-Planck equation. Surprisingly, the corrections to (\ref{eq:MeanResponse}) are negligible, but this must not be interpreted as a general conclusion since it is highly dependent on the structure of the spectrum, hence on the details of the model. In the next section, we will see that the two-moment truncation is essential to well capture the out-of-equilibrium dynamics in mean-field models.

\section{Mean-Field theories}

\begin{figure*}
    \centering
    \includegraphics[width=\textwidth]{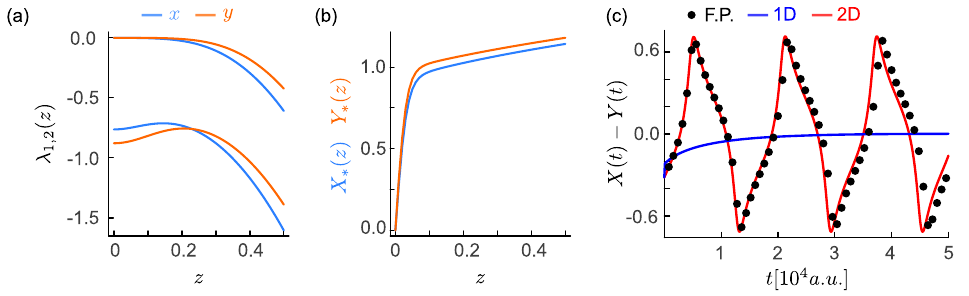}
    \caption{First two eigenvalues (a) and stationary mean (b) for the single $x$ variable (light blue) and the $y$ (orange) variable with respect to the input bias $z$. In figure (c), we match the coupled Fokker-Planck integration (black dots) with the one-dimensional moments dynamics (blue) and the two-dimensional moment dynamics (red). The parameters of the model are $a_e=0.9$, $a_i=1$, $\mu_{e,i}=0.1 a_{e,i}$, and $D=0.03$. }
    \label{fig:FPCoupled}
\end{figure*}

So far, we have dealt with non-linear systems described by linear operators $\mathcal{L}$. However, this may leave out the very important class of mean-field theory. To see this, one might consider a network of $N$ interacting processes each following:
$$dx_i = \left [f(x_i) +\frac{\sum_{j=1}^{N} x_j}{N} \right ]dt + \sigma dW_i $$
with $dW_i$ independent Wiener processes. If the number of processes is huge, $N \gg 1$, then the population mean approaches $X(t) \sim \frac{\sum_i x_i}{N}$, and the collective behavior is well described by:
$$dx=[f(x) +X(t)]dt + \sigma dW$$
The probability distribution $p(x,t)$ will then follow 
$$\partial_t p =\mathcal{L}_X p=  -\partial_x\left [f(x)+ X(t) \right]p +\frac{\sigma^2}{2}\partial_x^2p$$,
with the average self-consistently determined by $X(t) =\int_{\mathcal{D}} x p(x,t)dx$; hence, the operator is nonlinear. For a full account of the different kinds of nonlinear Fokker-Planck equations and their use, we refer to \cite{frank2005nonlinear}. The spectral expansion of this type of master equation takes the same form as (\ref{eq:ERE}), considering the mean field $X(t)$ as the parameter $\gamma$. This implies that the same benefits of using the moments system, introduced here for processes with time-dependent parameters, should apply to mean-field theories as well. As an example, consider the model introduced in \cite{gang1996stochastic} where a coupled excitatory population ($x_i$) and an inhibitory population ($y_i$) are coupled globally as follows:
\begin{equation}\label{eq:MFCoupled}
\begin{split}
    dx_i & = \left [a_e x_i -x_i^3  + \frac{\mu/a_e}{N}\sum_{j=1}^{N}(x_j-y_j) \right] dt + \sqrt{2D}dW_i \\
      dy_k & = \left [a_i y_k -y_k^3  + \frac{\mu/a_i}{N}\sum_{j=1}^{N}(x_j-y_j) \right] dt + \sqrt{2D}dW_k
\end{split}
\end{equation}
This kind of coupling is rather common and has been considered in several contexts \cite{lindner1995array}, ranging from neural networks \cite{amit1997model,brunel2000dynamics} to lasers \cite{jung1992collective}. In the mean-field limit $N \to \infty$, the system described by (\ref{eq:MFCoupled}) reduces to a two-dimensional process which is governed by two coupled Fokker-Planck equations (\ref{eq:FPCoupled}):
\begin{equation}\label{eq:FPCoupled}
\begin{split}
    \partial_t p_e(x,t) & = -\partial_x\left [a_e x -x^3  + \mu_e (X(t)-Y(t)) \right]p_e +D\partial_x^2p_e\\
      \partial_tp_i(y,t) & = -\partial_y \left [a_i y -y^3  + \mu_i(X(t)-Y(t)) \right]p_i + D\partial_y^2 p_i
\end{split}
\end{equation}
where the averages are self-consistently estimated:
$$X(t) =\int_{\mathcal{D}} x p_e(x)dx \hspace{1cm} Y(t) =\int_{\mathcal{D}} y p_i(y)dy$$
To apply moment projections, we first need to compute the eigenvalues of the two Fokker-Planck equations (\ref{eq:FPCoupled}), $\lambda_n^{(x)}$ and $\lambda_n^{(y)}$ respectively, as a function of the constant bias in the drift terms $z=\mu_{i,e}[X(t)-Y(t)]$. Due to the symmetry of the drift, $\lambda_n(-z)=\lambda_n(z)$, and the same goes for the asymptotic moments:
$$X_*(z) =\int_{\mathcal{D}} x \phi_0(x,z)dx \hspace{1cm} Y_*(z) =\int_{\mathcal{D}} y \phi_0(y,z)$$
The asymptotic moments and first two eigenvalues are shown in Fig. \ref{fig:FPCoupled} (a) and (b). The one-moment approximation ($G_{\lambda}=\lambda$) leads to:
\begin{equation}\label{eq:1DApprox}
    \begin{split}
        \dot{X}(t) &=\lambda_1^{x}[\mu_e Z(t)](X(t)-X_*(\mu_e Z(t))) \\
        \dot{Y}(t) &=\lambda_1^{y}[\mu_i Z(t)](Y(t)-Y_*(\mu_i Z(t)))
    \end{split}
\end{equation}
where $Z(t)=X(t)-Y(t)$. In \cite{gang1996stochastic}, the authors have shown that whenever:
$$\mu_{e,i}> \frac{a_{e,i}D(\lambda_1^{(x)}+\lambda_1^{(y)})}{\lambda_1^{(x)}-\lambda_1^{(y)}}$$
the solution $X=Y=0$ becomes unstable, leading to a limit cycle, as we report in Fig. \ref{fig:FPCoupled} (c). In this case, (\ref{eq:1DApprox}) fails to reproduce the expected behavior, but we can solve this problem by extending the dimensionality using our result (\ref{eq:G2m}). Introducing the expected values of the drift terms:
\begin{equation*}
    \begin{split}
        F(t) &=\int_{\mathcal{D}} (a_e x -x^3 +\mu_e Z(t) ) p_e(x,t)dx \\
        G(t) &=\int_{\mathcal{D}} (a_i y -y^3 +\mu_i Z(t) ) p_i(y,t)dy
    \end{split}
\end{equation*}
we can write the coupled two-moment system:
\begin{widetext}
\begin{equation}\label{eq:2DApprox}
%\small
\begin{pmatrix}
   \dot{X} \\ \dot{F} \\ \dot{Y} \\ \dot{G} \\
\end{pmatrix}
=    \begin{pmatrix}
0 & 1 &0 &0\\
 -\lambda_1^{(x)}(t) \lambda_2^{(x)}(t) & \lambda_1^{(x)}(t) +\lambda_2^{(x)}(t) & 0 & 0\\
 0& 0 & 0&1 \\
 0&0& -\lambda_1^{(y)}(t)\lambda_2^{(y)}(t) & \lambda_1^{(y)}(t) +\lambda_2^{(y)}(t)
\end{pmatrix} \begin{pmatrix}
X(t)-X_*(t)\\ F(t) \\Y(t)-Y_*(t) \\ G(t) \\
\end{pmatrix}
\end{equation}
\end{widetext}
where $\lambda_n$ and the asymptotic moments depend on time through $Z(t)$ as in (\ref{eq:1DApprox}). As shown in Fig. \ref{fig:FPCoupled} (c), system (\ref{eq:2DApprox}) not only captures the instability of the stationary solution but also reproduces quantitatively the limit cycle dynamics obtained from direct integration of (\ref{eq:FPCoupled}).
Note that even though in this paper we have mostly studied models where detailed balance is conserved, this example together with our recent work \cite{MattiaVinci2024} shows that our approach works also for non-equilibrium systems.

\section{Conclusions}

In this work, we have shown how to recast the standard spectral expansion approach, applied to master equations for probability density, to a set of ODEs for an arbitrary ensemble of observables, i.e. moments. 
Our approach is equivalent to standard spectral decomposition but allows greater analytical tractability due to a number of general constraints. 
Furthermore, we highlight the deep relation between the results of our work and the general theory of the Koopman operator and its data-driven estimation \cite{williams2015data}. 
We discuss the limitations of considering the moment system instead of the pure spectral decomposition, when the observables chosen are suboptimal, and present analytical constraints that might be used as a guiding principle for the optimization of optimal observables in data-driven approaches.
We demonstrate how to use the moment-projection technique to tackle the moment-closure problem using as a benchmark the case of the Bessel process with constant drift, where standard approximations such as Gaussian closure are not applicable. 
Moreover, in this case, the standard spectral expansion, despite full knowledge of eigenfunctions and eigenvalues, cannot be fully closed analytically in the open system case, i.e., when the parameters of the Fokker-Planck equation depend on time. 
On the contrary, the truncated moment system is fully determined at any order of the truncation, showing a remarkable agreement between the analytical low-dimensional moments system and detailed Monte Carlo simulations. 
On a more general level, our approach highlights the profound relation between the relaxation dynamics of the moments and the time scales of the system, i.e., the eigenvalues of the Fokker-Planck operator. 
Most importantly, it demonstrates that the moments dynamics of an open system, i.e., one driven by an external force, can be determined by looking exclusively at the relaxation dynamics of the unperturbed system, in the same spirit as the generalized fluctuation-dissipation theory \cite{Vulpiani2009-co}.
As another example, we apply our framework to the study of stochastic resonance and derive several general expressions for the spectral gain, some of which are, to the best of our knowledge, new. 
Finally, we demonstrate that our approach can be used as it is even for rather general mean-field models and it is capable of capturing at a quantitative level out-of-equilibrium dynamics. 
Several future directions are possible. 
In the companion paper \cite{MattiaVinci2024}, we introduce and succesfully exploited this theoretical framework to the challenging case of a mean-field theory for the dynamics of spiking neurons, an out-of-equilibrium system described by a nonlinear Fokker-Planck equation. 
It is then interesting to understand if the same can be done for other systems of interest where standard moment-closure techniques fail. 
Another important aspect that we have largely neglected here is the effect of the continuous component of the spectrum of eigenvalues; much work in this direction might be needed, and in this regard, it would be interesting to apply novel techniques that have been introduced recently in the context of the Koopman operator \cite{lusch2018deep}. 
In this work, we have tried to bridge the gap between different approaches that all aim ultimately to find a low-dimensional description of high-dimensional complex behaviors. 
Given the paramount importance of moment-closure in Science and the recent renewed interest sparked by data-driven and machine learning approaches, we believe that our results have the potential to both synthesize and generate new insights in this field.

\begin{acknowledgments}
Work partially funded by the NextGenerationEU and MUR (PNRR-M4C2I1.3) project MNESYS (PE0000006–DD 1553 11.10.2022) and project EBRAINS-Italy (IR0000011– DD 101 16.6.2022) to MM.
\end{acknowledgments}

\appendix

\section{Relation between $\vec{a}$ and $\vec{m}$}

The central result of this paper is the relation between eigenmodes $a_n(t)$ and moments $m_k(t)$ (\ref{eq:Trick}). We will derive it here for the most general case of both a continuous and discrete spectrum. In this case, the formal identity of the spectral expansion is:
$$p(x,t)=\phi_0(x) +\sum_{n=1}^{N} a_n(t)\phi_n(x)  +\phi_{c}(x,t)$$
where the continuous component eigenfunction can be written as:
$$\phi_c(x,t)=\int e^{\lambda t} g(\lambda,x) d\lambda$$
where the contour of the integral and the function $g(\lambda,x)$ depend on the specific details of the problem. We now define $\tilde{m}_{k,0}(t) = \int_{\mathcal{D}} x^k(\phi_0(x)+\phi_c(x,t))dx $ and the moment projection matrix as $U_{kn}=\int_{\mathcal{D}} x^k \phi_n(x)dx$. Multiplying by $x^k$ and integrating over $x$, we get an equation for each $k$:
$$m_k(t)-\tilde{m}_{k,0}(t)=\sum_{n}^{N}a_n(t) U_{kn}$$
which in vectorial notation leads to:
$$\vec{m}(t) -\vec{\tilde{m}}_0(t) = \mathbf{U} \vec{a}(t)$$
This relation is exact in the limit where we include all the $N=\infty$ discrete eigenmodes and only approximate otherwise. If the matrix $\mathbf{U}$ is invertible for the choice of observables $m_k$, we finally arrive at (\ref{eq:Trick}).

\section{Derivation of equation (\ref{eq:SR}) }

In the regime of linear response theory, we assume that the time-dependent driving, parameterized by a small amplitude $\epsilon$ via $\epsilon(t)$, is small ($\epsilon \ll 1$). Then, for any function $g(m(t))$, we can expand to first order $g(\vec{m}(t))\approx g(\vec{m}_*^0) +\epsilon \nabla g|_{\vec{m}_*^0} \cdot \vec{m}_1(t) +o(\epsilon^2)$, where $\vec{m}(t) = \vec{m}_*^0 + \epsilon \vec{m}_1(t)$ and $\vec{m}_*^0$ is the stationary state for $\epsilon=0$. The superscript $^{0}$ indicates the stationary component ($\epsilon=0$). Consider now the system \ref{eq:MainResult} with initial condition $\vec{m}(t=0)=\vec{m}_{*}^{0}$ (starting from equilibrium). In the Fourier domain (Laplace with $s=i\Omega$), linear response gives:
$$\vec{m}_1(\Omega) =(\mathbf{I}i\Omega -\mathbf{G^{0}})^{-1}\mathbf{G^{0}}\partial_{\epsilon}\vec{m}_*|_{\epsilon=0} $$
that leads to equation \ref{eq:AplificationFactor}.
The amplification factor $\eta$ of equation \ref{eq:SR} comes from the fact that, when only the first eigenvalue is considered, $G_{\lambda}=\lambda_1$ is used.
The same reasoning allows one to find the linear response in the form (\ref{eq:ERE}), see also \cite{MattiaVinci2024} % Changed placeholder to potentially relevant citation
for further details. In particular:
$$\vec{a}_1(\Omega)=\frac{i\Omega \vec{c}(\Omega)}{i\Omega\mathbf{I} - \mathbf{\Lambda}}$$ % Adjusted linear response for 'a'
$\mathbf{I}$ is the identity matrix. Finally, note that $m_1 = \partial_{\epsilon}m_*|_{\epsilon=0} + \mathbf{U} \vec{a}_1$.
Moreover, a 1D Fokker-Planck equation like the one in the text has a stationary solution of the Boltzmann type:
$$\phi_{0,\epsilon}(x)=e^{-\frac{\tilde{U}(x,\epsilon)}{D}}/Z(\epsilon)=e^{-\frac{U(x) + A(x)\epsilon}{D}}/Z(\epsilon)$$
where we also introduced the coefficient of normalization $Z(\epsilon)$ and the contribution from the external driving of order $\epsilon$.
For the general observable $m=\int_{\mathcal{D}}g(x,\epsilon)P(x,t)dx$:
\begin{equation*}
\begin{split}
\partial_\epsilon m_* |_{\epsilon=0} &=\partial_{\epsilon}\langle g(x, \epsilon) \rangle_{\epsilon}|_{\epsilon=0}=\frac{d}{d\epsilon}\int_{\mathcal{D}}g(x,\epsilon)\phi_0(x,\epsilon)dx |_{\epsilon=0} \\
                    &=\langle \partial_\epsilon g \rangle_0 -\langle g(x, 0) \rangle_{0}\frac{\partial_{\epsilon}Z|_{\epsilon=0}}{Z(0)} -\langle \frac{A(x)}{D} g(x,0) \rangle_{0}\\
                    &=\langle \partial_\epsilon g \rangle_0 + \frac{1}{D}( \langle \langle A(x) g(x,0) \rangle \rangle_{0}  ) % Using relation dZ/deps = -<A(x)>/D * Z
\end{split}
\end{equation*}
Finally, taking the limit $\epsilon \to 0$ and with the notation $\langle \langle A(x)B(x) \rangle \rangle =\langle A(x)B(x) \rangle - \langle A(x) \rangle \langle B(x)\rangle $, we arrive at equation \ref{eq:SR}, where the expectation values are with respect to the unperturbed $\phi_0$.

\section{Spectral quantities of the Bessel process with constant drift}

The study of the Bessel Process with constant drift can be found in numerous references, but we will refer to the results obtained in \cite{guarnieri2017}. The eigenvalue spectrum for this process is composed of a discrete and a continuous part. The discrete part is always dominant, and since we will truncate the spectral expansion to a relatively low number of modes, we can neglect the continuous part.
The eigenvalues are:
\begin{equation} \label{eq:BesselEig} % Added label for clarity
    \lambda_n =-\frac{2a^2 n b +a^2 n^2 \sigma^2}{2(b+n\sigma^2)^2}
\end{equation}

The eigenfunctions of the adjoint operator $\mathbf{L_{x}^{\dagger}}$ are:
\begin{equation}
    \psi_n(x) =Z_n (\xi_n x)^{b/\sigma^2} e^{-\xi_n x/2}L_{n}^{-1 +2b/\sigma^2}(\xi_n x)
\end{equation}
with $Z_n =\sqrt{\frac{\xi_n n!}{\Gamma(n+\frac{2b}{\sigma^2}) (2n+\frac{2b}{\sigma^2})}}$. The eigenfunctions of $\mathbf{L_x}$ are:
\begin{equation}
    \begin{split}
        \phi_0(x)=\frac{N}{\sigma^2} e^{-2 U(x)/\sigma^2} \\
        \phi_n(x)=e^{-U(x)/\sigma^2}\psi_{n}(x)
    \end{split}
\end{equation}
with the potential $U(x)=b \ln(x) +a x$ and the normalization coefficient
$$N=\left[ b 4^{-b/\sigma^2} \sigma^{4 b /\sigma^2 -2} (-a)^{-(2b/\sigma^2 +1)}\Gamma(\frac{2b}{\sigma^2}) \right]^{-1}$$
A convenient set of observables in this case are the standard moments: $m_k=\langle x^k |p \rangle$. The reason behind this is the polynomial nature of the drift term; in fact, the eigenfunctions are solutions of:
$$\lambda_n |\phi_n \rangle = -\partial_{x}(b/x +a)|\phi_n \rangle +\frac{\sigma^2}{2}\partial_{x}^2|\phi_n \rangle $$
applying $\langle x^k |$, we arrive at a hierarchy of equations for the entries of the projection matrix $U_{k,n}=\langle x^k |\phi_n\rangle$:
\begin{equation}\label{eq:constraints}
    U_{k,n}=-\frac{k(b U_{k,n} +a U_{k-1,n}) +\frac{\sigma^2 k (k-1)}{2} U_{k-2,n}}{\lambda_k}
\end{equation}
Since the normalization of $|p\rangle$ depends only on $|\phi_0\rangle$, we must have $U_{0,n}=0$. Moreover, if we choose $k>0$, thanks to (\ref{eq:constraints}), the only integrals we need to compute to close the moments system are $U_{-1,n}$ and $U_{k,0}$. If we define $\gamma=2a/\sigma^2$ and $\beta=2b/\sigma^2$, we find:
\begin{equation}
    \begin{split}
        & U_{k,0}=\frac{\Gamma(\beta +1 + k)(-\gamma)^k}{\Gamma(\beta +1)}\\
        &U_{-1,n}= (-\gamma +\xi_n)^{-n+\beta} (-\gamma -\xi_n)^n \sqrt{\frac{\xi_n ^{2\beta +1} \Gamma(n+\beta)}{2^{-2\beta}n!(2n+\beta)}} \\ % Check U_{-1,n} formula derivation if needed
    \end{split}
\end{equation}
The last result can be derived using the following identity:
\begin{equation*}
\begin{split}
    \int_{0}^{\infty}t^{\beta} e^{-st}L_{n}^{\gamma}(t) dt=\frac{\Gamma(\beta +1)\Gamma(\gamma+n+1)}{n! \Gamma(\gamma +1)}s^{-(\beta+1)}*\\
F(-n,\beta+1;\gamma+1;1/s) % Added semicolons for clarity in F arguments
\end{split}
\end{equation*}
valid for $\mathrm{Re}(s)>0$, and where $F(a,b;z)$ is the confluent hypergeometric function \cite{gradshteyn2014}.

\bibliography{VinciEtAlRefs}% Produces the bibliography via BibTeX.

\end{document}